\documentstyle[preprint,aps,tighten]{revtex}
\begin{document}
\draft
\input{epsf}

\title{Quasibosons}

\author{W. A. Perkins}
\address{(Perkins Advanced Computer Systems) Auburn, California }

\maketitle

\begin{abstract}
The similarity of the commutation relations for bosons 
and quasibosons (fermion pairs) suggests the possibility 
that all integral spin particles 
presently considered to be bosons could be quasibosons.
The boson commutation relations for integral spin particles
could be just an approximation to the quasiboson commutation
relations that contain an extra term.
Although the commutation relation for quasibosons are slightly
more complex, it is simpler picture of matter in that only
fermions and composite particles formed of fermions exist.
Mesons are usually referred to as
bosons, but they must be quasibosons since their internal structure 
is fermion (quark) pairs.
The photon is usually considered to be an elementary boson, but
as shown here, existing experiments do not rule out the possibility
that it is also a quasiboson. We consider how the quasiboson, composite nature
of such a photon might manifest itself.
\end{abstract}


%
\section{Introduction}
\label{sec.intro}

Most integral spin particles (light mesons, strange mesons, etc.) are composite particles formed of quarks. Because of their underlying fermion structure, these integral spin particles are not fundamental bosons, but composite quasibosons. However, in the asymptotic limit, which generally applies, they are essentially bosons. For these particles, Bose commutation relations are just an approximation, albeit a very good one. There are some differences; bringing two of these composite particles close together will force their identical fermions to jump to excited states because of the Pauli exclusion principle.

A few integral spin particles (photon, gluons, weak bosons, and gravitons) are regarded as elementary, exact bosons. For all of these particles, except the photon, there is no direct evidence from their statistics to differentiate between boson and quasiboson behavior. As we shall show in this paper, surprisingly, present experiments do not differentiate between a ``boson photon'' and a ``quasiboson photon.'' If the photon were a quasiboson, it would presumable be composed of neutrinos or massless quarks.

Since the predictions of quantum electrodynamics are in such excellent agreement with experiment (e.g., the calculated anomalous magnetic moment of the electron agrees to a few parts per billion with experiment), one might think there is little or no room for improvement in the model of the photon. In addition, since the photon is known to be massless to high precision, how can it be a composite particle? Would not a composite photon have readily detectable 
self-interactions? Does not the association of the photon with local gauge invariance indicate that it is elementary?

In spite of all its successes, there are some problems with the present photon model: Many of the calculations diverge (e.g., the calculated anomalous magnetic moment of the electron really gives infinity, before renormalization) and non-physical photon polarization states must be introduced to satisfy Lorentz invariance. Some of these problems are discussed in a recent paper concerning composite photons~\cite{perkins4}. A method of binding a fermion-antifermion pair with a zero range interaction that does not involve quanta is also discussed~\cite{perkins4}. With such an interaction the composite particle need not have mass or self-interactions. It is further shown~\cite{perkins4} that a composite-photon theory can be Lorentz invariant without the need for gauge invariance.

Let us compare the commutation relations for fermions, bosons, and quasibosons. Fermions are defined as the particles whose creation and
annihilation operators  obey the anticommutation relations 
\begin{eqnarray}
\{a({\bf k}),a({\bf l})\} = 0,  \nonumber \\
\{a^\dagger({\bf k}),a^\dagger({\bf l})\} = 0,  \nonumber \\
\{a({\bf k}),a^\dagger({\bf l})\} 
= \delta({\bf k}-{\bf l}),
\label{eqn1} 
\end{eqnarray}
while bosons are defined as the particles 
that obey the commutation relations,
\begin{eqnarray}
\left[b({\bf k}),b({\bf l})\right] = 0,  \nonumber \\
\left[b^\dagger({\bf k}),b^\dagger({\bf l})\right] = 0,  \nonumber \\
\left[b({\bf k}),b^\dagger({\bf l})\right] 
= \delta({\bf k}-{\bf l}).
\label{eqn2}
\end{eqnarray}

In superconductivity~\cite{blatt},
the Cooper pairs are referred to as
``quasibosons,'' since they obey commutation relations 
similar to, but different from, those of bosons. 
It is well known
that molecules with an even number of fermions 
exhibit Bose behavior, while
those with an odd number exhibit Fermi behavior~\cite{ehrenfest}. 
Theoretically, these composite
molecules formed of an even number of fermions 
(as well as nuclei formed of an
even number of fermions) do not obey 
Bose commutation relations so we will refer
to them as quasibosons.

The creation and
annihilation operators of quasibosons 
(composite particles formed of fermion pairs)
obey the commutation relations of the form,
\begin{eqnarray}
\left[Q({\bf k}),Q({\bf l})\right] = 0,  \nonumber \\
\left[Q^\dagger({\bf k}),Q^\dagger({\bf l})\right] = 0, \nonumber \\
\left[Q({\bf k}),Q^\dagger({\bf l})\right] 
= \delta({\bf k}-{\bf l})- \Delta({\bf k},{\bf l}).
\label{eqn3}
\end{eqnarray}

The commutation 
relations for a pair of fermions, Eq.~(\ref{eqn3}), are
similar to those of bosons, Eq.~(\ref{eqn2}). The 
$ \Delta( \bf k, \bf l )$ term (see Eq.~(\ref{eqn9})) 
looks complicated but its value is usually very small. Thus, it is
easy to envisage that Eq.~(\ref{eqn2}) is just an approximation
to Eq.~(\ref{eqn3}), 
the more accurate commutation relations for
integral spin particles.

As presented in many quantum mechanics texts it may appear that Bose 
statistics follow from basic principles, but it is really from 
the classical canonical formalism. This is not a reliable procedure
as evidenced by the fact that it gives the completely 
wrong result for spin-${1 \over 2}$ particles. Furthermore,
in extending the classical
canonical formalism for the photon, it is necessary to deviate
from the canonical rules (see Ref.~\cite{bjorken}, pp.~71,98).

Based on the symmetry of systems of identical
particles, it can be shown that their wavefunctions must be symmetric
or antisymmetric under interchange (see Ref.~\cite{bjorken}, pp.~32-34). 
Although identical 
bosons are symmetric under interchange, so are identical quasibosons.
It has also been claimed that the spin-statistics theorem
requires that integral spin particles must be bosons. 
In Sec.~5, 
the spin-statistics theorem is reviewed and
shown not to apply to quasibosons 
which do not satisfy space-like commutativity
because of their finite extent. 

In Sec.~6, the experimental evidence 
concerning the photon is
examined. We conclude that present experiments 
do not rule out the possibility that photons are quasibosons. 
Although the state of two elementary neutral bosons 
(which are their own antiparticle and
are identical) must be symmetric under exchange, two composite
quasibosons (which are not identical) can be antisymmetric
or symmetric under interchange.  
The author has suggested~\cite{perkins1} that there exist two
distinct $\pi^0$'s as a solution to the 
$\overline{p} p \rightarrow \pi^0\pi^0$ puzzle.
As discussed in Sec.~3, composite
photons can be non-identical also, making
possible axial vector meson decays into two photons (see
Sec.~6.2).

\section{Statistics of quasibosons}
\label{sec.stat}
\noindent
Consider quasibosons formed 
of two different types of fermions of equal mass
whose annihilation operators are given the symbols ``a'' and ``c''. 
Quasibosons, formed from pairs of spin-${1 \over 2}$ particles, 
have spin 0 or 1. Assuming that the
system is in a large box of finite volume 
with periodic boundary conditions, the
quasiboson annihilation 
and creation operators are defined 
as~\cite{lipkin,sahlin,landau1,perkins2},
\begin{eqnarray}
Q({\bf p}) = \sum_{\bf k} F^\dagger({\bf k})
c({\bf p}/2-{\bf k})a({\bf p}/2+{\bf k}), \nonumber \\
Q^\dagger({\bf p}) = \sum_{\bf k} F({\bf k})
a^\dagger({\bf p}/2+{\bf k})
c^\dagger({\bf p}/2-{\bf k}),
\label{eqn4}  \end{eqnarray}
where $ a( {\bf k} ) $ and $ c( {\bf k} ) $ are annihilation
operators for two different types of fermions.

The Fourier transform of the creation operator is,
\begin{equation}
Q^\dagger({\bf R}) = \int d{\bf r}\: \phi({\bf r})
\Psi_a^\dagger({\bf R}-{\bf r}/2)
\Psi_c^\dagger({\bf R}+{\bf r}/2),
\label{eqn5}
\end{equation}
where $ \phi({\bf r}) $ describes the relative 
motion and can be expanded in plane waves,
\begin{equation}
\phi({\bf r})  = \sum_{\bf k} F({\bf k})
e^{- i {\bf k} \cdot {\bf r}}.
\label{eqn6}
\end{equation}
The relative and center-of-mass coordinates 
and momenta are defined by,
\begin{eqnarray}
{\bf r} = {\bf r_1}-{\bf r_2},&
{\bf R} = {{\bf r_1}+{\bf r_2} \over 2}, \label{eqn7} \\
{\bf k} = {{\bf k_1}-{\bf k_2}\over 2}, &
{\bf p} = {\bf k_1}+{\bf k_2}, \nonumber
\end{eqnarray}  
with
\begin{equation}
\sum_{\bf k} \left| F({\bf k}) \right|^2 = 1.
\label{eqn8}
\end{equation}

The quasiboson operators of Eq.~(\ref{eqn4}) 
have been shown~\cite{lipkin,sahlin,landau1,perkins2},
to obey the commutation relations (\ref{eqn3}) with
\begin{eqnarray}
\Delta({\bf p}^{\prime},{\bf p}) = 
\sum_{\bf k} F^\dagger({\bf k}) \left[  
F({\bf p}^{\prime}/2-{\bf p}/2+{\bf k}) 
a^\dagger({\bf p}-{\bf p}^{\prime}/2-{\bf k})
a({\bf p}^{\prime}/2-{\bf k}) \right. \nonumber \\
\left.
+ \: F({\bf p}/2-{\bf p}^{\prime}/2+{\bf k}) 
c^\dagger({\bf p}-{\bf p}^{\prime}/2+{\bf k})
c({\bf p}^{\prime}/2+{\bf k}) \right]. 
\label{eqn9} 
\end{eqnarray}

For Cooper electron pairs~\cite{landau1},
``a'' and ``c'' represent different spin directions. For
nucleon pairs (the deuteron) ~\cite{lipkin,landau1}, 
``a'' and ``c'' represent proton and neutron. For
neutrino-antineutrino pairs~\cite{perkins2}, 
``a'' and ``c'' represent neutrino and antineutrino.
The size of the deviations from pure Bose behavior, 
$\Delta({\bf p}^{\prime},{\bf p}), $
depends on the degree
of overlap of the fermion wave functions 
and the constraints of the Pauli principle.

If we assume the state has the form,
\begin{equation}
|\Phi \rangle = a^\dagger({\bf k_1})
a^\dagger({\bf k_2})...a^\dagger({\bf k_n})
c^\dagger({\bf q_1})c^\dagger({\bf q_2})...c^\dagger({\bf q_m})|0 \rangle
\label{eqn9a}
\end{equation}       
then the expectation value of (\ref{eqn9}) vanishes for
${\bf p}^{\prime} \ne {\bf p}$, and we can approximate
the expression for $\Delta({\bf p}^{\prime},{\bf p})$ by,
\begin{eqnarray}
\Delta({\bf p}^{\prime},{\bf p}) = 
\delta({\bf p}^{\prime}-{\bf p})
\sum_{\bf k} \left| F({\bf k}) \right|^2 
\left[ a^\dagger({\bf p}/2-{\bf k})
a({\bf p}/2-{\bf k}) \right.  \nonumber \\  \left.
+ c^\dagger({\bf p}/2+{\bf k})
c({\bf p}/2+{\bf k}) \right].
\label{eqn10} 
\end{eqnarray}

Using the fermion number operators $n_a({\bf k}) $ and
$n_c({\bf k}) $, this can be written,
\begin{eqnarray}
\Delta({\bf p}^{\prime},{\bf p}) = 
\delta({\bf p}^{\prime}-{\bf p})
\sum_{\bf k} \left| F({\bf k}) \right|^2 
\left[ n_a( {\bf p}/2-{\bf k})
+ n_c({\bf p}/2+{\bf k})
\right] \nonumber \\
= \delta({\bf p}^{\prime}-{\bf p})
\sum_{\bf k} \left[ \left| F({\bf p}/2-{\bf k}) \right|^2 
 n_a({\bf k}) + \left| F({\bf k}- {\bf p}/2) \right|^2 
n_c({\bf k}) \right] \nonumber\\
= \delta({\bf p}^{\prime}-{\bf p})
\overline {\Delta} ({\bf p},{\bf p})
\label{eqn11} 
\end{eqnarray}
showing that it is the average number 
of fermions in a particular state ${\bf k}$ averaged
over all states with weighting factors 
$ F( {\bf p}/2-{\bf k}) $ and 
 $ F({\bf k}-{\bf p}/2) $.

The number operator for quasibosons is defined as,
\begin{equation}
N( {\bf p}) = Q^\dagger({\bf p}) Q({\bf p}).
\label{eqn12}
\end{equation}       
Using (\ref{eqn3}), (\ref{eqn9}), and (\ref{eqn12}), we obtain 
the following commutation relations for the
number operator,
\begin{eqnarray}
\left[ N({\bf p}^{\prime}),Q({\bf p}) \right]
=- \left \{ \delta({\bf p}^{\prime}-{\bf p})
-\Delta({\bf p}^{\prime},{\bf p}) \right \} 
Q( {\bf  p}^{\prime}), \nonumber \\
\left[ N({\bf p}^{\prime}),Q^\dagger({\bf p}) \right]
= Q^\dagger( {\bf  p}^{\prime})
 \left \{ \delta({\bf p}^{\prime}-{\bf p})
-\Delta({\bf p}^{\prime},{\bf p}) \right \}, \nonumber\\
\left[ N({\bf p}^{\prime}),N({\bf p}) \right] =  
Q^\dagger( {\bf p}) \Delta({\bf p}^{\prime},{\bf p}) 
 Q( {\bf  p}^{\prime})
- Q^\dagger( {\bf  p}^{\prime}) \Delta({\bf p}^{\prime},{\bf p}) 
 Q( {\bf  p}), \nonumber\\
\langle\left[ N({\bf p}^{\prime}),N({\bf p}) 
\right]\rangle \: = 0.
\label{eqn13}  
\end{eqnarray} 

As expected, these commutation relations differ 
from the usual Bose relations by
terms involving $\Delta({\bf p}^{\prime},{\bf p})$ . Note that 
$\Delta({\bf p}^{\prime},{\bf p})$  does not commute with the 
quasiboson annihilation and creation operators. 
Using the second of (\ref{eqn11}), we obtain,
\begin{eqnarray}
\left[ \overline{\Delta}({\bf p}, {\bf p}),
Q^\dagger({\bf q}) \right]
= \sum_{\bf k} \left \{ \left| F({\bf p}/2- 
{\bf k}) \right|^2
 \left[n_a( {\bf k}),
Q^\dagger( {\bf  q}) \right] \right.\nonumber \\ \left.
 + \left| F({\bf k} -{\bf p}/2) \right|^2
 \left[n_c( {\bf k}),
Q^\dagger( {\bf  q}) \right] \right \}. 
\label{eqn14}  
\end{eqnarray} 
Inserting (\ref{eqn4}) for $Q^\dagger({\bf q})$ and using the usual 
commutation relations for the fermion
number operator (see Ref.~\cite{landau1}, p. 456) gives,
\begin{eqnarray}
\left[ \overline{\Delta}({\bf p}, {\bf p}),
Q^\dagger({\bf q}) \right]
= \sum_{\bf k} \left \{ 
\left| F({\bf p}/2- {\bf q}/2 -{\bf k}) \right|^2 
+ \left| F({\bf q}/2-{\bf p}/2-{\bf k}) \right|^2
 \right \} \nonumber \\ \times \; F({\bf k}) a^\dagger({\bf q}/2+{\bf k})
c^\dagger({\bf q}/2-{\bf k}). 
\label{eqn15}
\end{eqnarray} 

At this point we need an approximation 
to obtain a workable value. Lipkin~\cite{lipkin} 
suggested for a rough estimate to assume that $F({\bf k})$  
is a constant for the states
used to construct the wave packet. He used the symbol $\Omega$ 
for the number of states
used to construct the wave packet. In that case, 
Eq.~(\ref{eqn8}) gives 
$ \left| F({\bf k}) \right|^2 = 1 / \Omega$, and
one obtains directly from (\ref{eqn15}),
\begin{equation}
\left[ \overline{\Delta}({\bf p}, {\bf p}),
Q^\dagger({\bf q}) \right]
= 2 \: Q^\dagger( {\bf q}) / \Omega({\bf p}, {\bf q}).
\label{eqn16}
\end{equation} 

In Lipkin's approximation, $\Omega$ does not depend 
upon ${\bf p}$ and ${\bf q}$.  An improvement can
be made by using,
\begin{eqnarray}
{1 \over \Omega({\bf p}, {\bf q})}
= {\sum_{\bf k} F({\bf k})  \left \{ 
\left| F({\bf p}/2- {\bf q}/2 -{\bf k}) \right|^2
+ \left| F({\bf q}/2-{\bf p}/2-{\bf k}) \right|^2
 \right \} \over {2 \: \sum_{\bf k} F({\bf k}) }}, 
\label{eqn17}
\end{eqnarray}
and letting  $F({\bf k})$  
be a Gaussian distribution, which satisfies the normalization 
condition (\ref{eqn8}),
\begin{equation}
F({\bf k}) = {(8 \pi)^{3/4} \over \sqrt{ V k_0^3}}
e^{-k^2/k_0^2},
\label{eqn18}
\end{equation}
where $V$ is the confinement volume. 
Going from a box to the infinite domain, Eq.~(\ref{eqn17}) becomes,
\begin{equation}
{1 \over \Omega({\bf p}, {\bf q})}
= {\int d^3 k \: F({\bf k}) 
\left| F({\bf p}/2- {\bf q}/2 -{\bf k}) \right|^2
\over \int d^3 k \: F({\bf k})  }. 
\label{eqn19}
\end{equation}
Inserting the Gaussian distribution for  $F({\bf k})$  
 and evaluating the integrals results in,
\begin{equation}
\Omega({\bf p}, {\bf q})
= \left({3 \over {8 \pi}}\right)^{3/2} V 
\left({ k_0 \over {\hbar c} }\right)^3
e^{ ({\bf p}-{\bf q})^2 / (6 k_0^2)}.
\label{eqn20}
\end{equation}

We can see from (\ref{eqn17}) that $1 / \Omega({\bf p},{\bf q})$  
will be very small if  $F({\bf k })$ and 
$F({\bf p}/2-{\bf q}/2-{\bf k})$  have little overlap. 
This can occur if ${\bf q} = -{\bf p}$ 
(two quasibosons emitted in opposite
directions) and $|{\bf p}|>>k_0$, the Gaussian width 
or momentum spread. This overlap
factor is given explicitly in the exponential of (\ref{eqn20}). 
For the case in which ${\bf q} = {\bf p}$,
 $\Omega({\bf p},{\bf q})$ does not depend upon ${\bf p}$, 
so we can just use $\Omega$. 

We can now use the second of 
(\ref{eqn13}) and (\ref{eqn16}) 
to find the effect of the quasiboson
number operator acting on a state of $m$ quasibosons,
\begin{equation}
N({\bf p}) (Q^\dagger({\bf p}))^m|0\rangle \;
= \left( m - {m(m-1) \over \Omega({\bf p}, {\bf p})}
\right) (Q^\dagger({\bf p}))^m|0\rangle,
\label{eqn22}
\end{equation}
where we have used 
$N({\bf p})|0\rangle \; = \: \overline{\Delta}({\bf p}, 
{\bf p})|0\rangle \; = \:0$. 
This result differs from the usual
one because of the second term which is small for large $\Omega$. 
Normalizing in the
usual manner (see Ref.~\cite{koltun}, p. 7),
\begin{eqnarray}
 Q^\dagger({\bf p})|n_{\bf p} \rangle \;
= \sqrt{ (n_{\bf p} +1) 
\left( 1- {n_{\bf p} \over \Omega} \right) }
|n_{\bf p} +1\rangle,  \nonumber \\
Q({\bf p})|n_{\bf p} \rangle \;
= \sqrt{ n_{\bf p} 
\left( 1- {(n_{\bf p}-1) \over \Omega} \right) }
|n_{\bf p} -1\rangle,
\label{eqn23}  
\end{eqnarray}
where $|n_{\bf p}\rangle$ is the state of $n_{\bf p}$ 
quasibosons having momentum ${\bf p}$ which is created
by applying $Q^\dagger({\bf p})$  on the vacuum $n_{\bf p}$ times. 
Note that,
\begin{eqnarray}
Q^\dagger({\bf p})|0 \rangle \; = \: | 1_{\bf p}\rangle, \nonumber \\
Q({\bf p})|1_{\bf p}\rangle \; = \: |0\rangle, 
\label{eqn24}
\end{eqnarray}
which is the same result as obtained 
with boson (or fermion) operators. In Eq.~(\ref{eqn23}) we see 
formulas similar to the usual ones with correction factors 
that approach zero for large $\Omega$. 

Now, let's look at possible commutation relations 
for a spin-1 quasiboson photon. The
quasiboson annihilation and creation operators 
need to be modified slightly to
handle spin and a mass zero composite particle. 
We consider photons to be composed of 
two-component neutrinos and their
antiparticle (momenta antiparallel and spins parallel). 
The neutrino and antineutrino are assumed to have 
antiparallel momentum when created and
absorbed in interactions with other particles. 
The photon is continuously creating
virtual pairs that annihilate. 
(Whether this is the correct composition for a real
photon is not of concern here. 
Our main interest is to have a model 
of a quasiboson photon for comparison purposes.) 
Let $\gamma_R({\bf p})$  
and $\gamma_L({\bf p})$  be annihilation
operators for right and left circularly polarized photons. Then,
\begin{eqnarray}
\gamma_R({\bf p}) = {1 \over \sqrt{2}}\sum_{\bf k} F^\dagger(k,{\bf n})
\left[ \: c_1(k,-{\bf n})\: a_1( p+ k,{\bf n})
+ \: c_2(p+k,{\bf n})\: a_2( k,-{\bf n}) \right],  \nonumber \\
\gamma_L({\bf p}) = {1 \over \sqrt{2}}\sum_{\bf k} F^\dagger(k,{\bf n})
\left[ \: c_2(k,-{\bf n})\: a_2( p+ k,{\bf n})
+ \: c_1(p+k,{\bf n})\: a_1( k,-{\bf n})\right],  
\label{eqn25}
\end{eqnarray}
where $ {\bf n} = {\bf p} / |{\bf p}|  = {\bf k} / |{\bf k}| $ 
and ``a'' corresponds to the neutrino and ``c'', the
antineutrino while the subscripts refer to different two-component 
neutrinos.  Another complexity, multiple two-fermion states, 
has also been introduced in Eq.~(\ref{eqn25}).
With the same approximation as used in Eq.~(\ref{eqn10}),
the commutation relations for $\gamma_R({\bf p})$  
and $\gamma_L({\bf p})$ become,
\begin{eqnarray}
\left[ \gamma_R({\bf p}^{\prime}), 
\gamma_R({\bf p}) \right] = 0, \nonumber \\
\left[ \gamma_L({\bf p}^{\prime}), 
\gamma_L({\bf p}) \right] = 0, \nonumber \\
\left[ \gamma_R({\bf p}^{\prime}), 
\gamma_R^\dagger({\bf p}) \right]
= \delta( {\bf p}^{\prime} - {\bf p}) 
(1 -{\overline \Delta_{12}}({\bf p},{\bf p})),  \nonumber \\
\left[ \gamma_L({\bf p}^{\prime}), 
\gamma_L^\dagger({\bf p}) \right]
= \delta( {\bf p}^{\prime} - {\bf p}) 
(1 -{\overline \Delta_{21}}({\bf p},{\bf p})), \nonumber \\
\left[ \gamma_R({\bf p}^{\prime}), 
\gamma_L({\bf p}) \right] = 0, \nonumber \\
\left[ \gamma_R({\bf p}^{\prime}), 
\gamma_L^\dagger({\bf p}) \right] = 0, 
\label{eqn26}
\end{eqnarray}
where 
\begin{eqnarray}
{\overline \Delta_{12}}({\bf p},{\bf p}) = {1 \over 2}
\sum_{\bf k} \left| F(k,{\bf n}) \right|^2 \left[  
a_1^\dagger(p+k, {\bf n})a_1(p+k, {\bf n}) 
+ \: c_1^\dagger(k, -{\bf n}) c_1(k, -{\bf n}) \right. \nonumber \\
\left. + c_2^\dagger(p+k, {\bf n}) c_2(p+k, {\bf n})
+ \: a_2^\dagger(k, -{\bf n}) a_2(k, -{\bf n}) \right].
\label{eqn27}
\end{eqnarray}

We now consider whether dissimilar 
quasibosons commute or
anticommute. In general, two unlike quasibosons will commute 
as pairs of non-identical fermions commute. 
However, if two dissimilar quasibosons contain the
same fermion(s), then they will commute 
with commutation relations similar to
Eq.~(\ref{eqn3}) with the  $\Delta({\bf k},{\bf l})$  
term modified by some appropriate factor.

\section{Symmetry of quasibosons under interchange}
\label{sec.symmetry}
\noindent
First we consider a state of two identical quasibosons.
Assuming that the system is 
in a large box of finite volume with periodic
boundary conditions, a state of two quasibosons in the 
formalism of Eq.~(\ref{eqn4}) is represented by,
\begin{equation}
|Q_1 Q_2\rangle \; = \sum_{ {\bf p_1},{\bf p_2}} 
f( {\bf p_1},{\bf p_2} ) Q^\dagger({\bf p_1}) 
Q^\dagger({\bf p_2}) |0\rangle.
\label{eqn29}
\end{equation}
Since the quasiboson creation operators commute, 
the state of two quasibosons
as given by Eq.~(\ref{eqn29}) is symmetric 
under interchange of the quasibosons.

While the state of two identical bosons must be 
symmetric under interchange, a state of non-identical bosons 
(or quasibosons)
can be either symmetric or antisymmetric. For example, a state 
of two identical $\pi^0$'s must be a state with even relative
orbital angular momentum, while a state of $\pi^+\pi^-$ can have
even or odd relative orbital angular momentum.

We will consider the symmetry of a state of two quasiboson photons.
Linearly polarized photon annihilation operators can be constructed from 
circularly polarized operators by,
\begin{eqnarray}
\xi({\bf p}) = {1 \over \sqrt{2}}\left[ \gamma_L({\bf p}) +
\gamma_R({\bf p}) \right], \nonumber \\
\eta({\bf p}) = {i \over \sqrt{2}}\left[ \gamma_L({\bf p}) -
\gamma_R({\bf p}) \right].
\label{eqn28a}
\end{eqnarray}
Using Eq.~(\ref{eqn25}) we obtain,
\begin{eqnarray}
\xi({\bf p}) = {1 \over 2}\sum_{\bf k} F^\dagger(k,{\bf n})
\left[ \: c_1(k,-{\bf n})\: a_1( p+ k,{\bf n}) 
+ \: c_2(p+k,{\bf n})\: a_2( k,-{\bf n}) \right. \nonumber \\
\left. + \: c_2(k,-{\bf n})\: a_2( p+ k,{\bf n}) 
+ \: c_1(p+k,{\bf n})\: a_1( k,-{\bf n}) \right],  \nonumber \\
\eta({\bf p}) = {i \over 2}\sum_{\bf k} F^\dagger(k,{\bf n})
\left[ \: c_1(k,-{\bf n})\: a_1( p+ k,{\bf n}) 
+ \: c_2(p+k,{\bf n})\: a_2( k,-{\bf n}) \right. \nonumber \\
\left. - \: c_2(k,-{\bf n})\: a_2( p+ k,{\bf n}) 
- \: c_1(p+k,{\bf n})\: a_1( k,-{\bf n})\right].  
\label{eqn29a}
\end{eqnarray}
From Eq.~(\ref{eqn29a}) we see that a state 
of two composite photons such as,
\begin{equation}
|\Phi\rangle \; = \sum_{ {\bf p_1},{\bf p_2}} 
f( {\bf p_1},{\bf p_2} ) \xi^\dagger({\bf p_1}) 
\eta^\dagger({\bf p_2}) |0\rangle.
\label{eqn30a}
\end{equation}
need not be symmetric under interchange as the two photons are not
identical.

\section{Commutation relations for quasiboson fields}
\label{sec.fields}
\noindent
Consider the field 
\begin{equation}
\phi( x) = A(x) + A^\dagger(x),
\label{eqn44}
\end{equation}
where
\begin{equation}
A( x) = {1 \over (2 \pi)^3 } 
\int d^3 p \: { 1 \over \sqrt{ 2 p_0}} Q( {\bf p})
\: e^{i p \cdot x},
\label{eqn45}
\end{equation}
with $ p \cdot x = {\bf p} \cdot {\bf x} - p_0 x_0 $. 
We will follow the usual argument, which shows that Bose
fields are local (e.g., see Ref.~\cite{veltman}, p. 25), except that 
Eq.~(\ref{eqn45}) has a quasiboson
operator $ Q( {\bf p} ) $ instead of a Bose operator. 
Using Eqs.~(\ref{eqn3}) and (\ref{eqn11}) for the 
quasiboson commutation relations gives, 
\begin{equation}
\left[ A( x), A^\dagger(y) \right] = {1 \over (2 \pi)^3 } 
\int d^3 p \: { 1 \over \sqrt{ 2 p_0}}
\: e^{i p \cdot (x-y)} \left[ 1 - \overline {\Delta} 
( {\bf p}, {\bf p}) \right].
\label{eqn46}
\end{equation}
Since
\begin{equation}
\left[ A^\dagger( x), A(y) \right] = 
- \left[ A( y), A^\dagger(x) \right],
\label{eqn46.5}
\end{equation}
Eqs.~(\ref{eqn44}) and (\ref{eqn46}) result in,
\begin{eqnarray}
\left[ \phi( x), \phi(y) \right] = 
{1 \over (2 \pi)^3 } 
\int d^3 p \: { 1 \over 2 p_0}
\: \left\{ e^{i p \cdot (x-y)} - \: e^{-i p \cdot (x-y)} 
\right\}  \left[ 1 - \overline {\Delta} 
( {\bf p}, {\bf p}) \right],
\label{eqn47}
\end{eqnarray}
which can be expressed as
\begin{eqnarray}
\left[ \phi( x), \phi(y) \right] = i D(x - y) 
- {i \over (2 \pi)^3 } 
\int d^3 p \: { 1 \over p_0}
\: \sin \left[ p \cdot (x-y) \right] 
 \overline {\Delta} ( {\bf p}, {\bf p}). 
\label{eqn48}
\end{eqnarray}

For equal times $ D( {\bf x}- {\bf y}, 0) $ is zero, 
but due to the second term this commutator
does {\it not} vanish for space-like intervals $ (x-y)^2 < 0 $. 
This means that quasibosons, being composite particles, 
have a finite extent in space. Quasiboson fields
cannot have a commutator which vanishes 
everywhere outside the light cone.
Otherwise, one could prove using the 
spin-statistics theorem that quasibosons are
bosons, a contradiction.

This field satisfies the Klein-Gordon equation 
and therefore might be appropriate
for a spin-0 composite particle. The non-local field effect 
carries over to quasibosons with spin, of course. 

The commutation relations for the electromagnetic fields 
of quasibosons in terms
of $  \overline {\Delta}_{12} ( {\bf p}, {\bf p}) $ and
$  \overline {\Delta}_{21} ( {\bf p}, {\bf p}) $  of Eq.~(\ref{eqn27}) 
are (see Appendix A of Ref.~\cite{perkins3}),
\begin{eqnarray}
\left[ E_i( x), E_j(y) \right] = \left( \delta_{ij} 
{\partial \over \partial x_0} 
{\partial \over \partial y_0}
- {\partial \over \partial x_i} 
{\partial \over \partial y_j} \right)
\cr \left\{ i D(x-y) - {i \over 16 \pi^3} \int d^3 p \: p_0^{-1}
\: \sin \left[ p \cdot (x-y) \right] 
\left( \overline {\Delta}_{12} ( {\bf p}, {\bf p}) +
\overline {\Delta}_{21} ( {\bf p}, {\bf p})  \right) 
\right\} \nonumber\\
\cr - {i \over 16 \pi^3} 
{ \partial \over \partial y_0}
\sum_{k=1}^3 \epsilon_{ijk} 
{ \partial \over \partial x_k}
\int d^3 p \: p_0^{-1}
\: \cos \left[ p \cdot (x-y) \right]
\left( \overline {\Delta}_{12} ( {\bf p}, {\bf p}) -
\overline {\Delta}_{21} ( {\bf p}, {\bf p})  \right), 
\label{eqn49a} 
\end{eqnarray}
\begin{equation}
\left[ H_i( x), H_j(y) \right] =
\left[ E_i( x), E_j(y) \right],
\label{eqn49b}
\end{equation}
and
\begin{eqnarray}
\left[ E_i( x), H_j(y) \right] =
-{ \partial \over \partial y_0}
\sum_{k=1}^3 \epsilon_{ijk} 
{ \partial \over \partial x_k} \nonumber \\
\cr \left\{ i D(x-y)
 - {i \over 16 \pi^3} \int d^3 p \: p_0^{-1}
\: \sin \left[ p \cdot (x-y) \right]
\left( \overline {\Delta}_{12} ( {\bf p}, {\bf p}) +
\overline {\Delta}_{21} ( {\bf p}, {\bf p})  \right) 
\right\} \nonumber\\
\cr - {i \over 16 \pi^3} 
\left( \delta_{ij} 
{\partial \over \partial x_0} 
{\partial \over \partial y_0}
- {\partial \over \partial x_i} 
{\partial \over \partial y_j} \right) \nonumber \\
\int d^3 p \: p_0^{-1}
\: \cos \left[ p \cdot (x-y) \right] 
\left( \overline {\Delta}_{12} ( {\bf p}, {\bf p}) -
\overline {\Delta}_{21} ( {\bf p}, {\bf p})  \right). 
\label{eqn49c} 
\end{eqnarray}

These quasiboson commutation relations differ 
from the usual photon
commutation relations because of the extra terms involving 
 $  \overline {\Delta}_{12} ( {\bf p}, {\bf p}) $ and
$  \overline {\Delta}_{21} ( {\bf p}, {\bf p}) $.
To estimate the deviation from local commutativity, we note that 
$  \overline {\Delta}( {\bf p}, {\bf p}) $  is
independent of $ {\bf p}$ in Lipkin's approximation 
(see Eq.~(6.13) of Ref.~\cite{lipkin}, so it can be
taken out from under the integral sign. With that change, 
the ``$\sin$'' terms in Eqs.~(\ref{eqn48}), 
(\ref{eqn49a}), and (\ref{eqn49c}) 
contain the factor $ D(x-y)$ 
which vanishes for space-like intervals.
The ``$\cos$'' terms fall off as $1 / |{\bf x} - {\bf y}|^2 $
for large ${\bf x} - {\bf y} $
for a mass-zero particle (see
Ref.~\cite{bjorken}, p. 171), and the factor 
$  \overline {\Delta}_{12} ( {\bf p}, {\bf p}) -
 \overline {\Delta}_{21} ( {\bf p}, {\bf p}) $ should be small 
as it vanishes
for equal numbers of right-handed and left-handed photons.

The departure from local commutativity allows 
an interference between a particle
created at ${\bf x}$ and one created at ${\bf y}$, 
but does not restrict the measurability of $ \phi( {\bf x} )$
or $ \phi^\dagger( {\bf x} )$. Similarly, the small interference 
indicated by Eqs.~(\ref{eqn49a}), (\ref{eqn49b}), 
and (\ref{eqn49c}) should not
significantly effect the measurability 
of the fields as long as we do not attach
physical meaning to the measurement 
of the field strength at a point, but to
averages over finite space-time regions~\cite{bohr}. 

\section{Spin-statistics theorem}
\label{sec.spinstat}
\noindent
In his 1940 paper, Pauli~\cite{pauli} concludes: 
``For integral spin the quantization according to the
exclusion principle is not possible.'' 
If we apply this to quasibosons, which have integral spin, the theorem 
simply states that quasibosons cannot obey fermi statistics.

However, one of the basic assumptions of the theorem, 
space-like commutativity, is also not satisfied by
composite integral spin particles 
since they do not obey Bose commutation relations. 
Therefore, this theorem does not apply 
to most of the known integral spin
particles (nuclei and molecules 
with an even number of fermions).

Although the spin-statistics theorem does not apply 
to composite integral spin
particles, Ehrenfest's and Oppenheimer's~\cite{ehrenfest} 
{\it approximate} rule does apply. It states
that composite particles formed of an even number 
of fermions obey Bose statistics
while those formed of an odd number 
of fermions obey Fermi statistics. If the
spins of the fermions are collinear, the predictions 
of this rule are the same as
those of the spin-statistics theorem. 
This rule is also supported by the second
quantization formalism as an even number 
of fermion creation operators commute
and an odd number anticommute, 
and the deviations caused by extra terms in the
commutation relations are small for tightly bound, 
well separated particles~\cite{sahlin}.

If integral spin ``elementary particles'' are 
formed of multiple fermions, the
theorem would also not apply to them. 
As Messiah and Greenberg~\cite{messiah} and von
Baeyer~\cite{baeyer} noted long ago, 
experimental tests are necessary to determine the
symmetry of ``elementary particles,'' without recourse
to the spin-statistics theorem.

\section{Comparison with experiment}
\label{sec.expr}
\noindent
Here we look at the quasiboson theory 
for the photon and the
corresponding experimental results. But before that, 
we will briefly consider $He^4$ and Cooper pairs,
{\it known} quasibosons. 

\subsection{Known quasibosons}
\label{sec.known}
\noindent
Even for the superfluid 
state of $He^4$ the molecules are well 
separated~\cite{huang} as the interatomic distances 
are $4 \times 10^{-8}$  cm while the hard sphere radius is
$1 \times 10^{-8}$  cm. Treating $He^4$ as a
boson is a good approximation 
as shown by many observables, such as specific
heat, ultrasound absorption, 
and neutron scattering. The Bose-Einstein
condensation is also in general agreement 
with this expectation. However, the
Helium ground state potential shows a short-range 
hard-core repulsion~\cite{feltgen} which
is believe to be caused by the Pauli principle.

Since the electron-electron
distances for Cooper pairs in a superconductor 
are about $10^{-8}$ cm and the pair size is about 
$10^{-4}$  cm, one might expect
that we could not put many quasibosons in the same state. 
On the contrary, the
Fermi statistics of the components 
does not prevent us from putting large
numbers of fermions pairs into one quasiboson state. 
As the density of 
quasibosons increases, the function, $F({\bf k})$  
spreads in momentum space to allow more
quasibosons in the same state. 
In superconductors, a Bose-Einstein 
like condensation occurs with a large number of pairs 
ending up in the
lowest energy state.
Experimental evidence that the Cooper pairs 
are not bosons is shown by the
energy gap. As Lipkin noted~\cite{lipkin}, 
``The Pauli principle effect thus produces an
energy gap in the excitation spectrum. 
This effect is characteristic of overlapping
fermion pairs, and would be absent 
if the fermions behaved like simple bosons.''

\subsection{Photon}
\label{sec.photon}
\noindent
The main evidence indicating that photons 
are bosons comes from the Blackbody
radiation experiments which are 
in agreement with Planck's distribution. 
We will now calculate the photon distribution 
for Blackbody radiation using the
second quantization method~\cite{koltun}, 
but with a {\it quasiboson} photon. 
The atoms in the
walls of the cavity are taken to be 
a two-level system with photons emitted from
the upper level  $\beta $  and absorbed 
at the lower level $\alpha $. 
The transition probability for
emission of a photon when  $n_{\bf p} $ photons are present 
is enhanced,
\begin{equation}
w_{\alpha \beta}( n_{\bf p} + 1 \leftarrow n_{\bf p} )
= (n_{\bf p} + 1) \left( 1 - {n_{\bf p} \over \Omega}
\right) w_{\alpha \beta}( 1_{\bf p}  \leftarrow 0 ),
\label{eqn50}
\end{equation}
where we have used the first of (\ref{eqn23}). 
The absorption is enhanced also, but less
since we use the second of (\ref{eqn23}),
\begin{equation}
w_{ \beta \alpha}( n_{\bf p} - 1 \leftarrow n_{\bf p} )
= n_{\bf p} \left( 1 - {n_{\bf p}-1 \over \Omega}
\right) w_{ \beta \alpha}( 0 \leftarrow  1_{\bf p}  ).
\label{eqn51}
\end{equation}
Using the equality,
\begin{equation}
w_{ \beta \alpha}( 0 \leftarrow  1_{\bf p}   )
= w_{ \alpha \beta}(  1_{\bf p} \leftarrow 0 ),
\label{eqn52}
\end{equation}
of the transition rates~\cite{koltun}, 
Eqs.~(\ref{eqn50}) and (\ref{eqn51}) can be 
combined to give,
\begin{equation}
{w_{\alpha \beta}( n_{\bf p}+1 \leftarrow n_{\bf p} )
\over
w_{ \beta \alpha}( n_{\bf p} - 1 \leftarrow n_{\bf p} )}
= {(n_{\bf p}+1) \left( 1 - {n_{\bf p} \over \Omega}
\right) \over
n_{\bf p} \left( 1 - {(n_{\bf p}-1) \over \Omega} 
\right) }.
\label{eqn53}
\end{equation}

According to Boltzmann's distribution law, 
the probability of finding the system
with energy E is proportional to $ e^{-E/kT} $. 
Thus, the equilibrium between emission
and absorption requires that,
\begin{equation}
w_{\alpha \beta}( n_{\bf p}+1 \leftarrow n_{\bf p} )
e^{-E_{\beta} /kT} =
w_{ \beta \alpha}( n_{\bf p} - 1 \leftarrow n_{\bf p} )
e^{-E_{\alpha} /kT}, 
\label{eqn54}
\end{equation}
with the photon energy $ \omega_p = E_{\beta} -  E_{\alpha} $. 
Combining (\ref{eqn53}) and (\ref{eqn54}) results in,
\begin{equation}
n_{\bf p} = {2 \over u+(u+2)/ \Omega +
\sqrt{u^2(1+2/\Omega) + (u+2)^2 /\Omega^2}},
\label{eqn55}
\end{equation}
with $u =  e^{\omega_p/kT} - 1$. 
For $\Omega(\omega_p/kT) >> 1$, this reduces to 
\begin{equation}
n_{\bf p} = {1 \over e^{\omega_p /kT}
\left( 1 + {1 \over \Omega} \right) - 1}.
\label{eqn56}
\end{equation}

For large $\Omega$  this approaches Planck's distribution law. 
The measured quantity in
Blackbody radiation experiments is usually $W_{\lambda}$, 
the spectral emittance as a function of wavelength,
\begin{equation}
W_{\lambda} = {2 \pi h c^2 \over \lambda^5
\left( e^{hc / \lambda kT}
\left( 1 + {1 \over \Omega} \right) - 1 \right) }.
\label{eqn57}
\end{equation}
The biggest deviations from Planck's law will occur for 
$\Omega h c /(\lambda k T) < 1$,  and in that
case Eq.~(\ref{eqn55}) must be used. 

We can calculate $\Omega( {\bf p},{\bf p})$ using Eq.~(\ref{eqn20}), 
but we need a value for the momentum
spread, $k_0$. From the uncertainty principle, 
the momentum spread must be of
order $ \hbar / \Delta x $ where $ \Delta x $  
is the photon wavelength,
\begin{equation}
k_0 \sim {\hbar \over \lambda}.
\label{eqn58}
\end{equation}
Using (\ref{eqn58}), Eq.~(\ref{eqn20}) becomes, 
\begin{equation}
\Omega({\bf p},{\bf p})=\left( {3 \over {8 \pi}}\right)^{3/2}
{V \over \lambda^3}.
\label{eqn20a}
\end{equation}

To see the effect of the $1 / \Omega( {\bf p},{\bf p})$  term, 
consider the Blackbody radiation
experiments of Coblentz~\cite{coblentz} 
at $1596^0$ $K$ in a cavity of 
volume 125 $cm^3$ in the
wavelength range 1 to 6.5 microns. For these conditions, 
Eq.~(\ref{eqn56}) applies 
and $1 / \Omega( {\bf p},{\bf p}) \le 10^{-9}$, 
and the maximum deviation from Planck's law is less than
one part in $10^{-8}$,
much too small to be detected. Comparison with other
Blackbody radiation experiments also showed that 
the $1 / \Omega( {\bf p},{\bf p})$  
term is too small to be detected.
Unfortunately, we cannot recommend any practical 
experimental test of Eq.~(\ref{eqn57}).

Another method of determining whether a particle is 
a quasiboson is by observing
the symmetry of two-particle states. As discussed in 
Sec.~3, the wave functions 
of two identical elementary bosons
must be symmetric under interchange 
while the wave functions of two composite quasiboson
can be antisymmetric if the two
quasibosons are not identical.

According to the theorem of Landau~\cite{landau2} and Yang~\cite{yang},
a vector particle (with total angular momentum $= 1$)
cannot decay into two photons. This can be seen as follows~\cite{close}:
The two photons state must be described in terms of three
vectors: the relative momentum ${\bf k}$, and the two polarization
vectors ${\bf \epsilon_1}$ and  ${\bf \epsilon_2}$. The
state must be bilinear in the polarization
vectors. There are just three possibilities:
\begin{eqnarray}
{\bf \epsilon_1} \times {\bf \epsilon_2}, \nonumber\\
({\bf \epsilon_1} \cdot {\bf \epsilon_2}) {\bf  k},  \nonumber\\
 {\bf  k} \times ({\bf \epsilon_1} \times {\bf \epsilon_2}).
\label{eqn51ab}
\end{eqnarray}
The last one has zero amplitude because of the transversality condition, 
 $ {\bf  k}\cdot {\bf \epsilon} = 0$. The first two are antisymmetric under an 
interchange of the two photons. Since two identical bosons or  two identical 
quasibosons are symmetric under interchange, a vector particle cannot decay into
two photons, completing the proof.

However, a vector particle can decay into two composite quasiboson photons, if 
the photons are not identical as in Eq.~(\ref{eqn30a}).
Many decays of vector particles into two photons are
forbidden by charge conjugation invariance. 
Since $n$ photons transform as $(-1)^n$, two photons 
will be even under charge conjugation. For example, the
$^3 S_1$,  and $^1 P_1$ states
of positronium have $C = -1$ and thus cannot decay into two photons. The 
 $^3 P_1$ state of positronium has $C = +1$ and can decay into two non-identical photons, providing a test of this theory.

Most of the vector mesons cannot decay into two photons because of charge conjugation
invariance, but some axial vector mesons such as $f_1(1285)$,
$f_1(1420)$, and $\chi_{c1}(1P)$ with $J^{PC} = 1^{++}$ can. Detection of such decays can provide
evidence that the photon is composite
particle with non-identical forms.

\section{Conclusions}
\label{sec.conc}
\noindent
It is our conjecture that all integral 
spin particles are quasibosons, composed of
fermions. This is based on the observation 
that most known integral-spin particles
are quasibosons which behave so similar 
to bosons that it is difficult to detect the
non-Bose effects caused by the underlying fermions. 
The experimental results
regarding the photon, which is usually held up 
as the exemplar boson, are
inconclusive. It was shown in Sec.~6.2 that 
the Blackbody radiation from quasiboson photons 
is so similar to Planck's distribution that 
the difference could not
have been detected in any existing experiment. 
It was shown in Sec.~4 and 5 
that the spin-statistics theorem does not apply to
composite particles because their fields are non-local. 

In this paper, we have only considered the photon for it is
usually considered to be the model boson. Tests to determine
whether two photons are always symmetric under interchange were
discussed in Sec.~6.2.

Although the commutation relations, Eqs.~(\ref{eqn3}) 
and (\ref{eqn9}), are more complex than those
for bosons, it is really a simpler picture 
of matter in that there exists only
fermions and composite particles formed of fermions. 
The old approximation is still
valid: If a composite particle 
is formed of an odd number of fermions, use Fermi
statistics, and if a composite particle 
is formed of an even number of fermions, use
Bose statistics. Of course, in some cases, 
such as Cooper pairs, this is not a good
approximation.

\acknowledgments

The author would like to express his thanks 
to Professor J. E. Kiskis of the
University of California at Davis for many helpful discussions.

\end{document}